\pdfoutput=1
\documentclass[11pt,twoside,a4paper,cmspaper,final,collab]{cms-tdr}
\def\svnVersion{c5d1de4}\def\svnDate{2023/07/28}\def\cmsCernNoTag{CERN-EP-2021-100}\def\cmsCernDate{\today}\def\cmsMessage{Published in Physics Letters B as \href{http://dx.doi.org/10.1016/j.physletb.2023.137905}{\doi{10.1016/j.physletb.2023.137905}.}}
\begin{document}\cmsNoteHeader{HIN-18-008}

\newcommand{\PP}{\ensuremath{{\Pp}{\Pp}}\xspace}
\newcommand{\pPb}{\ensuremath{{\Pp}\text{Pb}}\xspace}
\newcommand {\PbPb}  {\ensuremath{\text{PbPb}}\xspace}
\newcommand{\VnDelta}{\ensuremath{V_{n\Delta}}\xspace}
\newcommand{\VnDeltaTwo}{\ensuremath{V_{n\Delta}}\xspace}
\newcommand{\PhotonP}{\ensuremath{{\PGg\Pp}}\xspace}
\newcommand{\PhotonPb}{\ensuremath{{\PGg}\text{Pb}}\xspace}
\newcommand{\DeltaEtaF}{\ensuremath{{\Delta\eta^F}}\xspace}
\newcommand{\Ntrack}{\ensuremath{\mathrm{N_{\text{trk}}^{\text{offline}}}}\xspace}
\newcommand{\Ntrackoff}{\ensuremath{\mathrm{N_{\text{trk}}^{\text{off}}}}\xspace}
\newcommand{\Ntrackavg}{\ensuremath{\langle N_{\text{trk}}^{\text{offline}} \rangle}\xspace}
\newcommand{\Ntrig}{\ensuremath{N_\text{trig}}\xspace}
\newcommand{\Npair}{\ensuremath{N^\text{pair}}\xspace}
\newcommand{\VoneDelta}{\ensuremath{{V_{1\Delta}}}\xspace}
\newcommand{\VtwoDelta}{\ensuremath{{V_{2\Delta}}}\xspace}
\newcommand{\VthreeDelta}{\ensuremath{{V_{3\Delta}}}\xspace}
\newcommand{\vTwo}{\ensuremath{v_2}\xspace}
\newcommand{\vThree}{\ensuremath{v_3}\xspace}
\newcommand{\vN}{\ensuremath{v_{n}}\xspace}
\newcommand{\DeltaEta}{\ensuremath{{\Delta\eta}}\xspace}
\newcommand{\DeltaPhi}{\ensuremath{{\Delta\phi}}\xspace}
\newcommand{\pTassoc}{\ensuremath{{p\mathstrut}^{\text{assoc}}_{\mathrm{T}}}\xspace}
\newcommand{\pTtrig}{\ensuremath{{p\mathstrut}^{\text{\mathstrut{trig}}}_{\mathrm{T}}}\xspace}
\newcommand{\pT}{\ensuremath{p_{\mathrm{T}}}\xspace}
\newcommand{\MeanpT}{\ensuremath{\langle p_{\mathrm{T}} \rangle}\xspace}
\newcommand{\trigEff}{\ensuremath{\varepsilon_{\text{trig}}}\xspace}
\newcommand{\zvtx}{\ensuremath{z_{\mathrm{vtx}}}\xspace}
\providecommand{\cmsTable}[1]{\resizebox{\textwidth}{!}{#1}}

\cmsNoteHeader{HIN-18-008}  
\title{Two-particle azimuthal correlations in \texorpdfstring{\PhotonP}{gamma-p} interactions\\ using \texorpdfstring{$\pPb$}{pPb} collisions at \texorpdfstring{$\sqrtsNN = 8.16\TeV$}{sqrt(s_NN) = 8.16 TeV}}

\date{\today}

\abstract{The first measurements of the Fourier coefficients (\VnDelta) of the azimuthal distributions of charged hadrons emitted from photon-proton (\PhotonP) interactions at the LHC are presented. The data are extracted from 68.8\nbinv of ultra-peripheral proton-lead (\pPb) collisions at $\sqrtsNN = 8.16\TeV$ using the CMS detector. The high energy lead ions produce a flux of photons that can interact with the oncoming proton. This \PhotonP system provides a set of unique initial conditions with  multiplicity lower than in photon-lead collisions but comparable to recent electron-positron and electron-proton data. The \VnDelta coefficients are presented in ranges of event multiplicity and transverse momentum (\pt) and are compared to corresponding hadronic minimum bias \pPb results. For a given multiplicity range, the mean  \pt of charged particles is smaller in  \PhotonP than in \pPb collisions. For both the \PhotonP and \pPb samples, \VoneDelta is negative, \VtwoDelta is positive, and \VthreeDelta consistent with 0. For each multiplicity and \pt range, \VtwoDelta is larger for \PhotonP events. The \PhotonP data are consistent with model predictions that have no collective effects.}

\hypersetup{%
pdfauthor={CMS Collaboration},%
pdftitle={Two-particle azimuthal correlations in photon-proton interactions with the CMS experiment}, 
pdfsubject={CMS},%
pdfkeywords={CMS, physics, photon-proton, collectivity}}

\maketitle 

\section{Introduction}

A wide variety of measurements suggest the existence of collectivity in the collisions of small systems such as the  proton-proton (\PP) \cite{cms:ppfirst,Aad:2015gqa,cms:ppsecond,Khachatryan:2016txc,add_atlas_pp:01} and proton-nucleus (pA) \cite{cms:pPbfirst, add_rhic_pA:01,add_rhic_pA:02,Aad:2012gla,Aad:2013fja,Abelev:2012ola,Aaij:2015qcq,ABELEV:2013wsa,Khachatryan:2015waa,cms:pPbPbPb_corr_identifiedPar,Aaboud:2017acw,Aaboud:2017blb} collisions. Such collectivity could indicate the formation of a hot, strongly interacting ``quark gluon plasma'' (QGP), characterized by nearly ideal hydrodynamic behavior \cite{PhysRevD.46.229,Heinz:2013th,Gale:2013da}, or could alternatively arise from gluon saturation in the initial state ~\cite{Dusling:2015gta, Nagle:2018eea}. Properties of the QGP have been previously studied in a wide range of high-energy nucleus-nucleus (AA) collisions at the CERN LHC and BNL RHIC \cite{add_rhic_AA:01,add_rhic_AA:04,cms:PbPbfirst,cms:PbPbsecond,ALICE:2011svq,ATLAS:2012at,star:AAfirst,star:AAsecond,phenix:AAfirst,STAR:oct2019,STAR:oct2018}. In these studies, collectivity is observed via the azimuthal correlations of particles that are far apart in rapidity. This phenomenon is known as the ``ridge"~\cite{Dusling:2015gta}, and has been unexpectedly observed in high-multiplicity \PP and \pPb collisions since the start of the LHC operation \cite{cms:ppfirst,Aad:2015gqa,cms:ppsecond,Khachatryan:2016txc,add_atlas_pp:01, add_rhic_pA:01,add_rhic_pA:02,cms:pPbfirst,Aad:2012gla,Aad:2013fja,Abelev:2012ola,Aaij:2015qcq,ABELEV:2013wsa,Khachatryan:2015waa,cms:pPbPbPb_corr_identifiedPar,Aaboud:2017acw,Aaboud:2017blb}. The two-particle azimuthal correlations can be characterized by their Fourier components (\VnDeltaTwo) where $n$ represents the order of the moment. If the two-particle correlations can be factorized into the product of the corresponding single particle azimuthal distributions, then the single-particle azimuthal anisotropy Fourier coefficients \vN can be extracted as $\vN = \sqrt{\VnDelta}$ ~\cite{Voloshin:1996}. The second (\vTwo) and third (\vThree) coefficients are known as elliptic and triangular flow, respectively, and are directly related to the initial collision geometry and its fluctuations, which influence the medium evolution and provide information about its fundamental transport properties \cite{geometry_and_fluctuations_01,geometry_and_fluctuations_02,geometry_and_fluctuations_03,elliptical_triangular}. 

In  high-multiplicity events, \vTwo and \vThree depend upon the hadron species \cite{cms:pPbPbPb_corr_identifiedPar,cms:april2018,Sirunyan:2018nqr,Adam:2015vsf,add_rhic_AA:02,add_rhic_AA:03} 
and scale with the number of valence quarks in the hadron \cite{cms:pPbPbPb_corr_identifiedPar}. Such results suggest a common origin of the collectivity seen in \PbPb, as well as in high-multiplicity \PP and \pPb events, where a hydrodynamic description can be used to reasonably reproduce the measurements in each case ~\cite{Weller:2017tsr,Bozek:2011if,Bozek:2012gr,Bozek:2013uha}. Probing systems with even smaller interaction regions is therefore important to understand the reach of such a hydrodynamic description. The search for collectivity has been recently extended to electron-positron ($\mathrm{e^+e^-}$), electron-proton ($\mathrm{ep}$), photon-proton (\PhotonP), and photon-nucleus interactions~\cite{alepCorr:2019, Belle:2022fvl, zeus:Dec2019_ep, ZEUS:2021qzg, ATLAS:2021jhn}. So far, no long-range near-side ridge has been detected in these systems. In $\mathrm{e^+e^-}$ collisions~\cite{alepCorr:2019, Belle:2022fvl}, strong exclusion limits have been set on the ridge yield, while in $\mathrm{ep}$ collisions (deep inelastic scattering and photoproduction)~\cite{zeus:Dec2019_ep, ZEUS:2021qzg}, the extracted Fourier coefficients are finite but do not conclusively imply collective behavior. In photon-nucleus collisions~\cite{ATLAS:2021jhn}, finite \vTwo and \vThree are measured after applying a template fit procedure to remove noncollective correlations, assuming they scale with multiplicity.

High-energy \pPb ultra-peripheral collisions at the LHC, where the impact parameter is larger than the nucleus radius provide a new system to extend the search of long-range correlations to photon-proton collisions. At \TeV energies, the lead (Pb) nuclei generate a very large quasi-real photon flux \cite{cms:fsq_16_012_01}. In the equivalent photon approximation \cite{cms:fsq_16_012_02,cms:fsq_16_012_03,cms:fsq_16_012_04}, this flux can be considered as $\PGg$ beams of virtuality $Q^{2} < 1/R^2$, where $R$ is the effective radius of the charge distribution. For Pb nuclei at 2.56\TeV with radius $R \approx 7\unit{fm}$, the quasi-real photon beams have virtualities $Q^{2} < 10^{-3}\GeV^2$, but very large longitudinal energy, up to $E_{\PGg} = \hbar c/\alpha R \approx 73\GeV$, where $\alpha$ is the reciprocal Lorentz relativistic factor.

This study complements recent results from small collision systems, such as $\mathrm{e^+e^-}$ and $\mathrm{ep}$~\cite{alepCorr:2019, Belle:2022fvl, ZEUS:2021qzg}. The CMS detector has been used to collect a large sample of \PhotonP interactions that occur in ultra-peripheral \pPb collisions. The beam energies were 6.50\TeV for the protons and 2.56\TeV per nucleon for the Pb nuclei, resulting in a center-of-mass energy per nucleon pair ($\sqrtsNN$) of 8.16\TeV. The resulting \PhotonP center-of-mass energy can fluctuate up to $\sim$1.4 TeV. The \PhotonP results are compared to both hadronic minimum bias (MB) \pPb collisions (previously studied in Ref.~\cite{Sirunyan:2017uyl}) and predictions of the {\PYTHIA} v8.2 \cite{Helenius:2019gbd} model interfaced with the Delphes v3.4.2 fast simulation package \cite{deFavereau:2013fsa}. The minimum bias data are compared to predictions from the \HIJING v2.1 generator \cite{Wang:1991hta} coupled to a full \GEANTfour simulation of the detector \cite{Agostinelli:2002hh}. 

\section{Experimental apparatus and data sample}
 
The central feature of the CMS apparatus is a superconducting solenoid of 6 m internal diameter, providing a magnetic field of 3.8 T. Within the solenoid volume is the silicon tracker, a lead tungstate crystal electromagnetic calorimeter, and a brass and scintillator hadron calorimeter, each composed of a barrel and two endcap sections that cover the range $\abs{\eta}<3.0$. The silicon tracker measures charged particles within the range $\abs{\eta}<2.5$. It consists of 1440 silicon pixels and 15\,148 silicon strip detector modules, and provides an impact parameter resolution of about $15\mum$ and a transverse momentum (\pt) resolution better than 1.5\% at $\pt\approx100\GeVc$. Event selection for this analysis makes use of detectors in the forward region: hadron forward (HF) calorimeters that use quartz fibers embedded in a steel absorber covering the region $3.0 < \abs{\eta} < 5.2$ and the two Zero Degree Calorimeters (ZDCs) which measure neutral particles with $\abs{\eta}>8.3$ \cite{Suranyi:2021ssd}. Analysis in the midrapidity region is based upon objects produced by the CMS particle-flow (PF) algorithm~\cite{Sirunyan:2017ulk}, which reconstructs and identifies final-state particles with an optimized combination of information from the various elements of the CMS detector. A more detailed description of the CMS detector, together with a definition of the coordinate system used and the relevant kinematic variables, can be found in Ref.~\cite{Chatrchyan:2008zzk}.

The analysis is performed using data recorded by CMS during the LHC \pPb run in 2016 with an integrated luminosity of 68.8~\nbinv. The proton-going direction is towards the side of the detector with positive $\eta$. As a result of the energy difference between the colliding beams, the nucleon-nucleon (NN) center-of-mass for \pPb collisions is not at rest with respect to the laboratory frame. Massless particles emitted at $\eta_\text{cm} = 0$ in the NN center-of-mass frame will be detected at $\eta = \ensuremath{+}0.465$ in the laboratory frame. The event samples were collected by the CMS experiment with a two-level trigger system \cite{Sirunyan:2017uyl} consisting in the level-1 (L1), where events are selected by custom hardware processors and the high-level trigger (HLT), that uses fast versions of the offline software. 

Samples of both \PhotonP-enhanced and MB events were collected requiring energy deposits in at least one of the HF calorimeters above a threshold of approximately 1\GeV at L1. The HLT system requires the presence of at least one charged particle (track) with $\pt > 0.4\GeVc$ in the pixel tracker. Track reconstruction was performed online as part of the HLT trigger with a reconstruction algorithm that is identical to the one used offline \cite{highqualitytracks}. More details of the MB trigger can be found in Ref.~\cite{Khachatryan:2016bia}. For each event, the reconstructed vertex with the highest number of associated tracks was selected as the primary vertex. A zero bias trigger requiring only the presence of proton and lead bunches in the CMS detector was used to independently study the trigger efficiency (\trigEff). The beam bunches were detected by induction counters placed 175\unit{m} from the interaction point on each side of the experiment. In addition, a sample of events with neither beam present was collected for noise studies.

\section{Event selection}

\begin{figure*}[htb]
\centering
\includegraphics[width=0.89\textwidth]{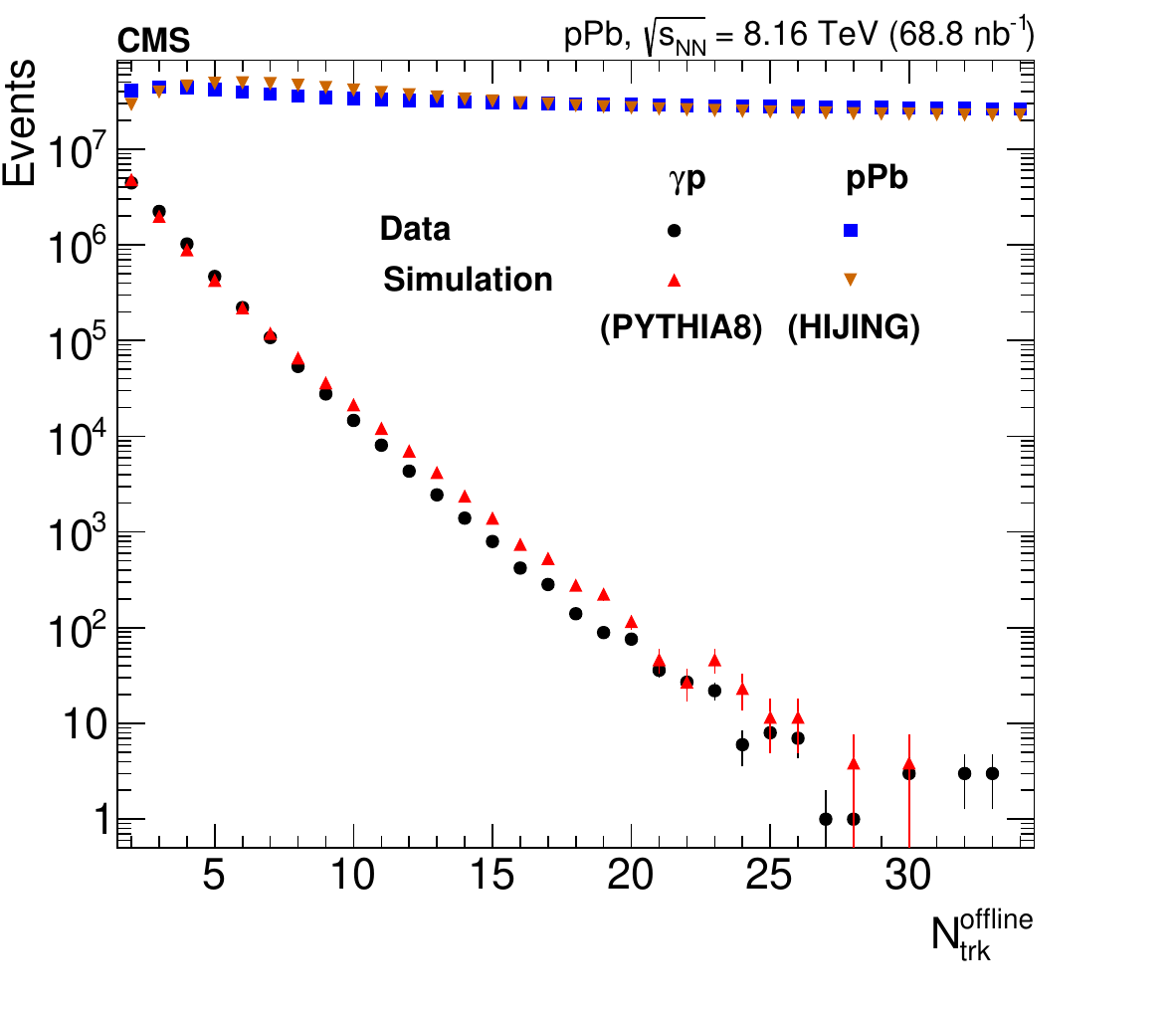}
\caption{The \Ntrack spectra for \PhotonP and MB samples. The simulated \PhotonP distribution has been normalized to the same event yield as the \PhotonP-enhanced data sample.}
\label{fig:spectra}
\end{figure*} 
 
For both \PhotonP and MB samples, the reconstructed primary vertex was required to be within 15\unit{cm} of the nominal interaction point along the beam axis ($z$) and within 0.15\unit{cm} in the transverse plane. The strategy for track selection is described in Ref.~\cite{highqualitytracks}. The impact parameter significance of reconstructed tracks with respect to the primary vertex in the longitudinal and transverse directions was required to be $<$3 standard deviations. Finally, the relative uncertainty in the \pt of the track was required to be $<$10\%. At least two reconstructed tracks with $\abs{\eta}<2.4$ and $\pt > 0.4\GeVc$ were required to be associated with the primary vertex. Beam-related background was suppressed by rejecting events for which $<$25\% of all reconstructed tracks pass the standard track selection criteria as in Ref.~\cite{Sirunyan:2017uyl}. 
 
Typical \pPb collisions produce particles at both positive and negative rapidity~\cite{Sirunyan:2017uyl,Sirunyan:2017vpr,Sirunyan:2018nqr}. However, \PhotonP events are expected to be very asymmetric in the laboratory frame since the photon energy is generally much smaller than the proton beam energy. 

For the \PhotonP-enhanced selection, a rapidity gap is defined as a continuous region in which there is low detector activity, as done in Ref.~\cite{HIN-18-019}. The detector acceptance $\abs{\eta}<5.0$ is divided into 20 bins. Threshold values are assigned to each $\eta$ bin, they delimit the energy from all PF candidates that can be considered significant and which contain at least 99.7\% of detector activity caused by detector noise or by beam-gas events. These thresholds were obtained by studying the zero-bias events triggered on noncolliding bunches. For each event, a given $\eta$ bin was considered to be empty if the energy registered from the PF candidates was below its assigned threshold value. For the 10 bins in the regions $\abs{\eta} < 2.5$ the energy threshold was 6\GeV and no high-purity tracks with $\pt>200\MeVc$ were allowed. For the four bins  from $-5.0 < \eta < -3.0$ in the lead-going region the thresholds were 16.9, 15.3, 16.4, and 13.4 \GeV, respectively. For the bin $-2.5 > \eta > -3.0$ only neutral hadrons were considered and the energy threshold  was 13.4\GeV. The forward rapidity gap (\DeltaEtaF) variable was then defined as the difference from $\eta=-5.0$ to the lower edge of the first nonempty $\eta$ bin. 

The MB selection requires the coincidence of at least one tower with energy above 3.0\GeV in both HF calorimeters and at least two tracks with $\abs{\eta}<2.5$. In contrast, a \PhotonP-enhanced selection is designed to capture events with an intact Pb nucleus, particle production in the positive $\eta$ region, and a large rapidity gap \cite{FRG:01,FRG:02,FRG:03}. The first two requirements are met by requiring no neutrons in the ZDC on the Pb-going side and at least 10\GeV in the highest energy tower of the HF calorimeter on the p-going side. To ensure a large rapidity gap, we require $5.0 < \DeltaEtaF < 7.5$. This corresponds to not having a particle within the negative-$\eta$ region. A total of $8.6\times10^6$ \PhotonP-enhanced and $1.0\times10^9$ MB candidate events were selected. In Ref.~\cite{HIN-18-019} the purity of the \PhotonP-enhanced sample with the ZDC selection is estimated to be about 95\% and it is weakly dependent on particle multiplicity. The requirement of no neutron emission used in this analysis gives an additional suppression of pomeron-Pb events. 

The reconstructed track multiplicity (\Ntrack) is defined as the number of tracks from the primary vertex with $\pt > 0.4\GeVc$, and $\abs{\eta}<2.4$. Figure~\ref{fig:spectra} shows the \Ntrack spectra for the \PhotonP-enhanced and MB data samples along with simulations from the {\PYTHIA}8 and \HIJING event generators. For the \PhotonP-simulated sample, the events are restricted to those with no tracks in the  $\eta < 0$ region and normalized to the \PhotonP-enhanced yield. In contrast to the MB sample, the \PhotonP-enhanced spectrum drops very rapidly with multiplicity up to a limiting value of 34. The \Ntrackavg value corresponding to the $2\le \Ntrack < 35$ range for the \PhotonP-enhanced sample is ${\approx}2.9$ and about 16.6 for the MB sample. The \Ntrack distribution from the zero bias data control sample has $\Ntrackavg \approx 0.84$. The \PhotonP-simulated sample shows a shape and range that is consistent with the \PhotonP-enhanced data sample. Three \Ntrack bins are used to analyze the \PhotonP-enhanced events: $2 \le \Ntrack < 5$, $5\le \Ntrack < 10$, $10\le \Ntrack < 35$. The first two deliver a comparable number of particle pairs and the third one aims to probe the higher \Ntrack domain by averaging the last part of the distribution. Table~\ref{tab:MeanMult} indicates the \Ntrackavg values for the data and simulated \PhotonP and MB samples. The mean \pt, \MeanpT, values of charged particles in the \PhotonP and MB data samples are $0.67 \pm 0.01$ and $0.74 \pm 0.01\GeVc$ respectively.
  
\begin{table*}
\centering
\topcaption{Mean \Ntrack for the \PhotonP-enhanced and the MB data sets for five classes of \Ntrack (abreviated as \Ntrackoff). Statistical uncertainties are negligible.}
\begin{tabular}{l c c c c c }  
\hline
Sample & $2 \leq\Ntrackoff < 5$ & $5\leq \Ntrackoff < 10$ & $10\leq \Ntrackoff < 35$ & $5\leq \Ntrackoff < 35$ & $2 \leq\Ntrackoff < 35$ \\ \hline
\PhotonP-enhanced  &  2.6 & 5.8 & 11.3  &   6.0   & 2.9  \\
\PhotonP-simulated &  2.6 & 5.9 & 11.4  &   6.2   & 2.9  \\
MB                 &  3.0 & 6.9 & 21.5  &   18.5  & 16.6 \\ 
MB-simulated       &  3.1 & 6.9 & 20.7  &   17.2  & 15.7 \\
\hline 
\end{tabular}
\label{tab:MeanMult} 
\end{table*}

\section{Analysis technique}

To ensure a high tracking efficiency, only tracks with $0.3 < \pt < 3.0\GeVc$ are used in the analysis. The two-particle correlation analysis techniques described below are identical to those used in previous CMS measurements in \PP, \pPb, and \PbPb collisions \cite{cms:ppsecond,cms:pPbfirst,cms:PbPbsecond}. For each multiplicity class, the ``trigger particles'' are tracks whose \pt, labeled as \pTtrig, is within a particular given range. The number of trigger particles in the event is denoted by \Ntrig. Particle pairs are then formed by associating each trigger particle with the remaining tracks whose \pt is denoted as \pTassoc. In this analysis \pTtrig and \pTassoc have a common range. Two different \pt ranges are studied, i.e., $[0.3, 3.0]$ and $[1.0, 3.0]\GeVc$. These are the same as those used in previous studies of the ridge \cite{cms:pPbfirst} and observations of correlations between \vN coefficients \cite{Sirunyan:2017uyl} in \pPb collisions.

The two-dimensional correlation function is defined as
\begin{linenomath*}
\begin{equation}
\frac{1}{\Ntrig}\frac{\rd^2\Npair}{\rd\DeltaEta \rd\DeltaPhi} = B(0,0)\frac{S(\DeltaEta, \DeltaPhi)}{B(\DeltaEta, \DeltaPhi)}, 
\end{equation}
\end{linenomath*}

\begin{figure*}[htbp]
\centering
\includegraphics[width=\textwidth]{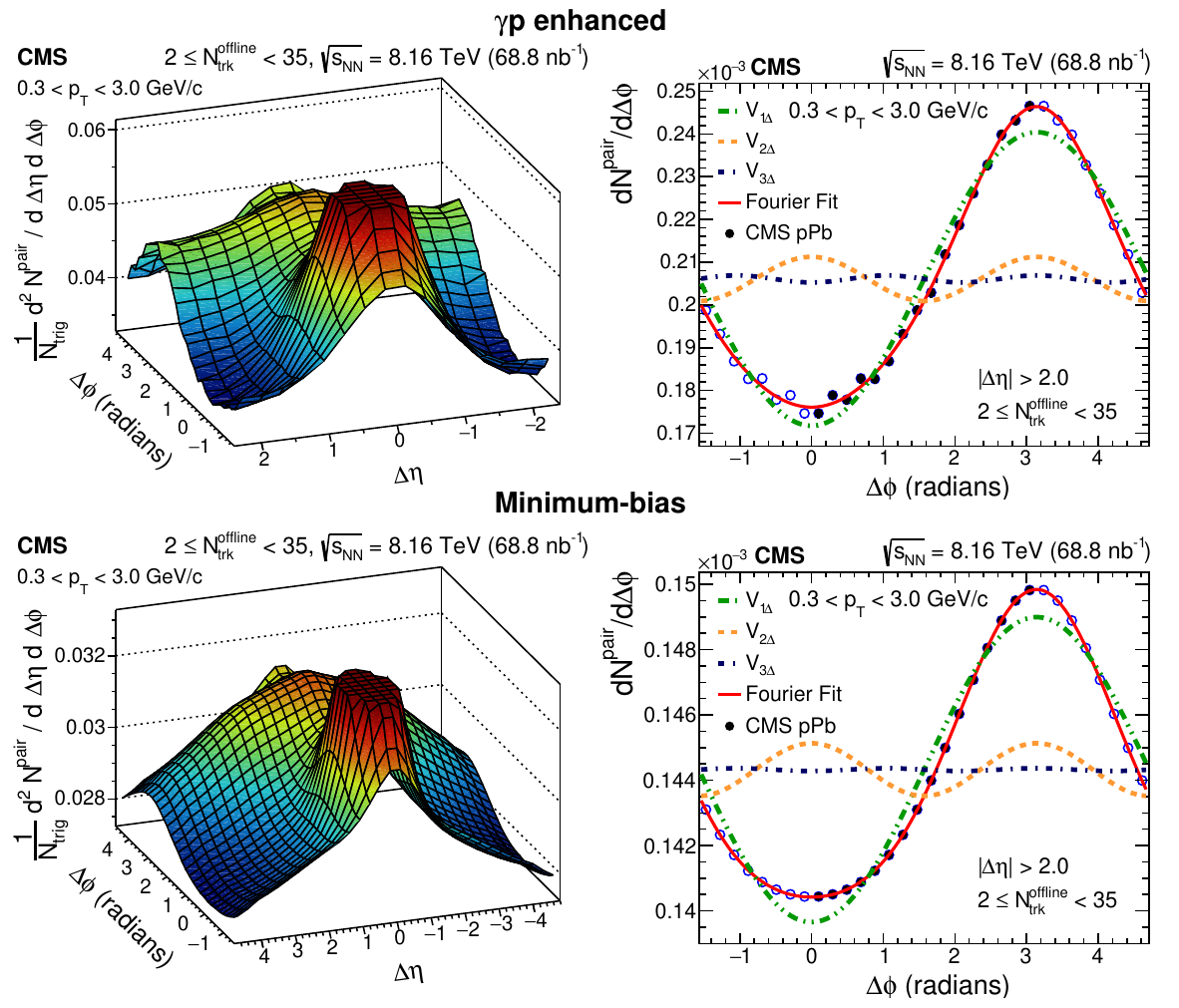}
\caption{Two-dimensional (left) and one-dimensional (right) correlation plots for \PhotonP-enhanced (upper) and MB (lower) events for  $0.3 < \pT < 3.0\GeVc$ and $2 \leq \Ntrack < 35$. For the two-dimensional distributions, the jet peak centered at $\DeltaEta = \DeltaPhi = 0$ is truncated to increase visibility. The rapidity gap requirement for the \PhotonP-enhanced sample limits the $\abs{\Delta\eta}$ range to $\abs{\Delta\eta} < 2.5$. The one-dimensional \DeltaPhi distributions are symmetrized by construction around $\DeltaPhi = 0$ and $\pi$. The Fourier coefficients, \VnDelta in the right column are fit over the $\DeltaPhi$ range $[0, \pi]$. Points outside this range are shown as open circles and are obtained by symmetrization of those in $[0, \pi]$. Statistical error bars are shown for both one-dimensional distributions.}
\label{fig:CorrelationPlots}
\end{figure*}

where \DeltaEta and \DeltaPhi are the differences in $\eta$ and $\phi$ of the pair and \Npair is the number of pairs. The same-event pair distribution, $S(\DeltaEta, \DeltaPhi)$, represents the yield of particle pairs from the same event in a given (\DeltaEta, \DeltaPhi) bin. Entries have been weighted by the product of inverse efficiencies evaluated for the kinematics of the two particles. The mixed-event pair distribution $B(\DeltaPhi, \DeltaEta)$ is constructed by pairing the trigger particles in each event with the associated charged particles from 100 different randomly selected events in the same 0.5 \cm wide vertex range and from the same track multiplicity class. It accounts for random combinatorial backgrounds and pair-acceptance effects. The same-event and mixed-event pair distributions are first calculated for each event, and then averaged over the events within the track multiplicity class. The mixed-event distribution is normalized by the sum of background events. The ratio $B(0,0)/B(\DeltaEta, \DeltaPhi)$ is the pair-acceptance correction factor, where $B(0,0)$ represents the mixed-event associated yield for both particles of the pair going in the same direction and thus having maximum pair acceptance.

Figure~\ref{fig:CorrelationPlots} (left) shows the two-particle correlation functions for \PhotonP-enhanced (upper row) and MB (lower row) events within the multiplicity range $2 \leq \Ntrack < 35$ as functions of \DeltaEta and \DeltaPhi. This \Ntrack range integrates all the yields all statistics for \PhotonP events, significantly suppressing fluctuations seen in smaller bins. For the \PhotonP distribution, the \DeltaEta range is limited to $\abs{\DeltaEta} < 2.5$ by the \DeltaEtaF selection and the acceptance of the tracker. Both distributions show a large jet peak centered at $\DeltaEta = \DeltaPhi = 0$, as well as a broader distribution from the recoiling jet centered at $\DeltaEta = 0$ and $\DeltaPhi = \pi$.  
Neither distribution displays a ``ridge''-like structure at $\abs{\Delta\phi} \approx 0$ for $\abs{\Delta\eta} > 2$. Figure~\ref{fig:CorrelationPlots} (right) shows the projections of the two-dimensional correlation functions onto the \DeltaPhi axis for $\abs{\DeltaEta} > 2$, away from the jet fragmentation peak. These distributions are fitted over the $\DeltaPhi$ range $[0, \pi]$ to a Fourier decomposition series $\propto 1 + \sum_n 2\VnDelta\cos(n\Delta\phi)$, from where the measured \VnDelta are extracted. Only the first three terms are included in the fit, since additional terms have a negligible effect on its quality.  

\begin{figure*}[htbp]
\includegraphics[width=\textwidth]{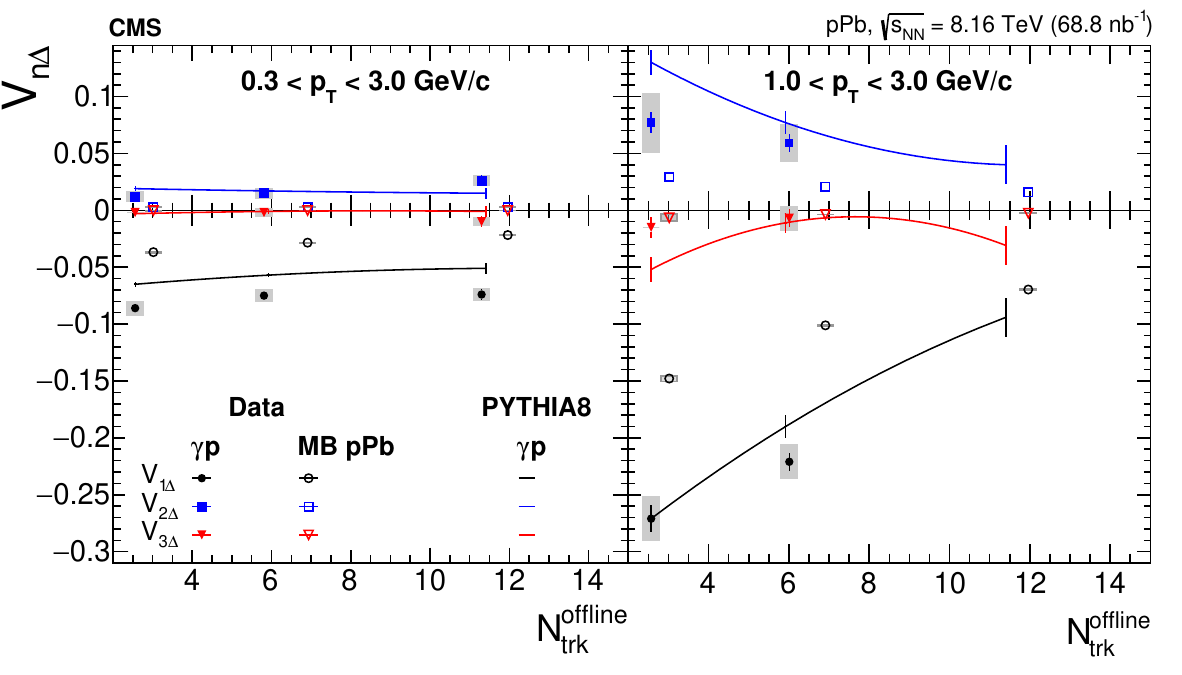}
\caption{Dependence of \VnDelta on \Ntrack for \PhotonP  and MB events for  two different \pt ranges. Systematic uncertainties are shown by the shaded bars in the two panels. The $2 \le \Ntrack < 5$, $5 \le \Ntrack < 10$, $10 \le \Ntrack < 35$ are used for the lower \pt range and $2 \leq \Ntrack < 5$ and $5 \leq \Ntrack < 35$ for the higher range. The points are placed at the mean value of the corresponding \Ntrack range. Lines indicate the prediction for \PhotonP events from {\PYTHIA}8.
}
\label{fig:MeasuredCoeffDependence}
\end{figure*}

In order to reduce the contribution to \vN coefficients from back-to-back jet correlations, one can correct \vN by subtracting correlations from very low-multiplicity events ($\ensuremath{v^\mathrm{sub}_n}$), as done in Refs. \cite{Khachatryan:2016txc, cms:pPbthird, Sirunyan:2017uyl}. The limited \Ntrack range for the \PhotonP events (up to $\sim35$) makes it difficult to apply this technique to check for collectivity. As an alternative the data are compared to {\PYTHIA}8 predictions, which do not include collective effects.

\section{Systematic uncertainties}

The systematic uncertainties of the experimental procedure are evaluated by varying the analysis conditions and extracting new \VnDelta coefficients. The following effects were considered:

\begin{enumerate} 
   \item The systematic uncertainties associated with the choice of the \DeltaEtaF range, which has a resolution of 0.5 units in $\eta$ and ensures low detector activity on one half of the detector, were estimated by repeating the analysis with $\DeltaEtaF \in [4.5, 5.0)$, just below the range of the nominal analysis. This alternative selection affects the track multiplicity and decreases the purity of the \PhotonP-enhanced sample up to 8\%~\cite{HIN-18-019}. The estimated size of this uncertainty has maximum values of 7\% for \VoneDelta and 27\% for \VtwoDelta within the \Ntrack range considered in this analysis. For the MB data there is no rapidity gap requirement and so no systematic uncertainty is assigned for this effect. 
   \item The effect of tracking inefficiency and misreconstructed track rate was studied by varying the track quality requirements. The selection thresholds on the significance of the transverse and longitudinal track impact parameter were varied from 2 to 5 standard deviations. In addition, the relative \pt uncertainty is varied from 0.05 to 0.10.  
   This translates into a 3.5\% uncertainty in \VoneDelta for the $2 \leq \Ntrack < 5$ category.
   \item The sensitivity of the results to the primary vertex position along the beam axis (\zvtx) was quantified by comparing events with different \zvtx locations from $-15$ to $+15\unit{cm}$. The magnitude of this systematic effect goes up to 150\% for \VthreeDelta with numerical estimations of ${\pm}0.003$  for $5 \leq \Ntrack < 10$ and $10 \leq \Ntrack < 35$ respectively, in the $0.3 < \pT < 3.0\GeVc$ category, and up to ${\pm}0.013$ for $1.0 < \pT < 3.0\GeVc$. 
   \item The trigger efficiency depends upon \Ntrack. It  decreases substantially for $\Ntrack < 10$, reaching  70\% for $\Ntrack = 2$. To study this effect, a parallel data sample with weighted events as (1/\trigEff) was produced. The full difference of the \VnDelta with and without the correction was taken as the uncertainty. This uncertainty is 2.3\% for \VoneDelta and 17\% for \VtwoDelta for  the sample with $2 \leq \Ntrack < 5$. 
\end{enumerate}

The systematic uncertainties were added in quadrature. For the \PhotonP-enhanced sample with $\Ntrack<35$ the final uncertainties in \VnDelta are 8.4 and 31\% for $n=1$ and 2, respectively. For the minimum bias sample the 
uncertainties for \VtwoDelta are 11\% for $2 \leq \Ntrack < 5$ and smaller than 2.6\% for the rest of the \Ntrack range.
Since $\pTtrig$ and $\pTassoc$ have the same range, the fractional uncertainties in \vN are half those of \VnDelta.
 
\begin{table*}[htbp]
\centering
\topcaption{The \VnDelta coefficients for \PhotonP-enhanced events, as functions of \pt and  \Ntrack. Statistical and systematic uncertainties are added in quadrature. 
} 
\begin{tabular}{lcccc} 
\hline 
\pt range  &  &  $2 \leq \Ntrack < 5$  &  $5 \leq\Ntrack < 10$ & $10 \leq \Ntrack < 35$\\
& \VoneDelta &  $-0.086 \pm 0.006$ &  $-0.075 \pm 0.005$ &  $-0.074 \pm 0.007$\\
$0.3 < \pT < 3.0\GeVc$      & \VtwoDelta  &  \hspace{0.32cm}$0.012 \pm 0.004$ &  \hspace{0.32cm}$0.015 \pm 0.004$ & \hspace{0.32cm}$0.026 \pm 0.006$\\
& \VthreeDelta &  $-0.002 \pm 0.001$ & $-0.002 \pm 0.004$ & $-0.010 \pm 0.006$\\
\hline
&  &  $2 \le \Ntrack < 5$  & \multicolumn{2}{c}{$5 \leq\Ntrack < 35$}\\
& \VoneDelta &  $-0.271 \pm 0.021$ & \multicolumn{2}{c}{ $-0.221 \pm 0.017$} \\
$1.0 < \pT < 3.0\GeVc$ & \VtwoDelta  &  \hspace{0.32cm}$0.077 \pm 0.027$ & \multicolumn{2}{c}{\hspace{0.32cm}$0.059 \pm 0.017$} \\
& \VthreeDelta &  $-0.015 \pm 0.009$ &\multicolumn{2}{c}{$-0.007 \pm 0.013$} \\
\hline
\end{tabular} 
\label{tab:results}
\end{table*}
 
\section{Results}

Figure~\ref{fig:MeasuredCoeffDependence} and Table~\ref{tab:results} show the measured \VnDeltaTwo coefficients as a function of \Ntrack for the two different \pT ranges for the \PhotonP and MB \pPb samples. For the MB sample, the results  are consistent with those in \cite{Sirunyan:2017uyl} before the subtraction procedure. Both the \PhotonP and MB distributions show a negative \VoneDelta, a positive \VtwoDelta of smaller magnitude than \VoneDelta, and a \VthreeDelta that is consistent with zero. For a given \Ntrack and \pt range, both \VoneDelta and \VtwoDelta are larger in the \PhotonP samples than in the MB results. For both samples, the magnitude of \VoneDelta tends to decrease with \Ntrack, while \VtwoDelta has at most a weak \Ntrack dependence. Their magnitudes are both larger in the higher \pt range. 

\begin{figure*}[htbp]
\includegraphics[width=\textwidth]{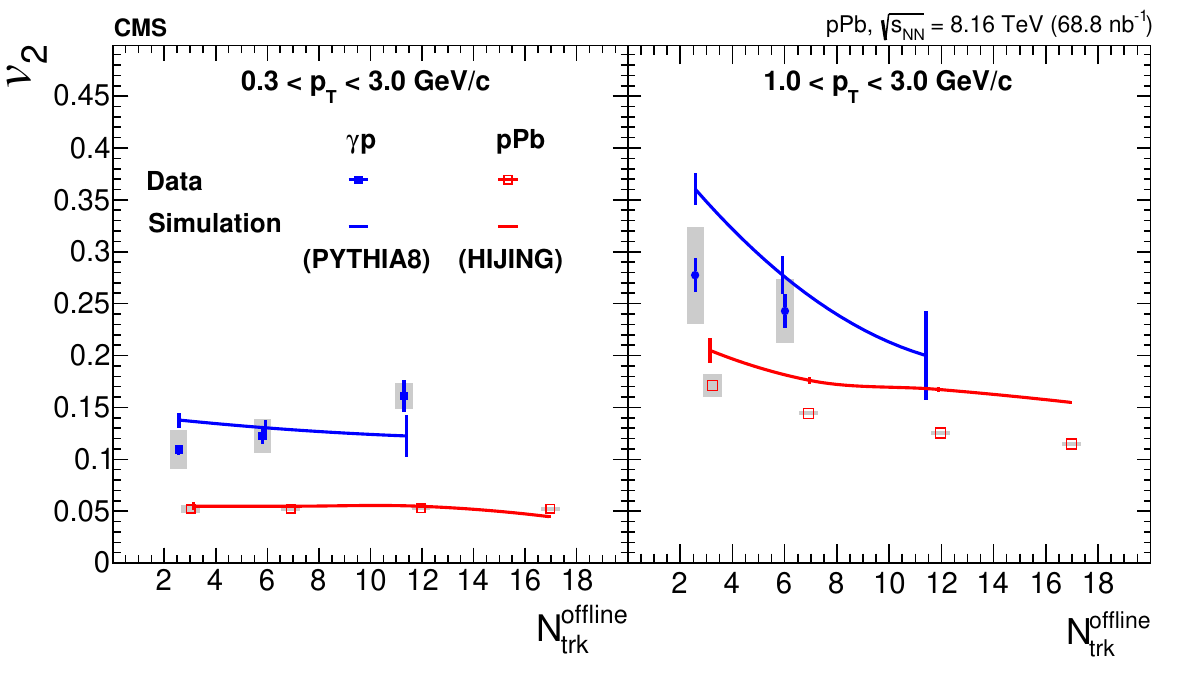}
\caption{\label{fig:v2Ntrack} Single-particle azimuthal anisotropy \vTwo versus \Ntrack for \PhotonP-enhanced and \pPb samples in two \pt regions. Systematic uncertainties are shown by the shaded bars in the two panels. 
Predictions from the {\PYTHIA}8  and \HIJING  generators are shown for the \PhotonP and MB \pPb samples  respectively. For the \PhotonP events, same \Ntrack bin arrangement as in Figure 3 is kept while for \pPb the bins $[2, 5)$, $[5, 10)$, $[10, 15)$ and $[15, 20)$ are used.} 
\end{figure*}

Figure~\ref{fig:MeasuredCoeffDependence} also shows predictions from the {\PYTHIA}8 generator for \VnDeltaTwo from \PhotonP collisions. The predictions of \VtwoDelta and \VthreeDelta from {\PYTHIA}8 are reasonably consistent with the \PhotonP data and have a similar dependence upon \pT and \Ntrack. The \VoneDelta prediction is smaller in magnitude than the measured values for the low \pT range. 

Figure~\ref{fig:v2Ntrack} shows \vTwo as a function of \Ntrack  and \pt for both \PhotonP and MB data sets. For $0.3 < \pT < 3.0\GeVc$, the MB results are consistent with previously published CMS results \cite{Sirunyan:2017uyl}. Predictions from the {\PYTHIA}8 and \HIJING  generators  are also shown for \PhotonP and MB \pPb interactions respectively, none of the models include collective effects. For both data and simulations, \vTwo varies slowly with track multiplicity for the \PhotonP and \pPb samples. At a given \Ntrack, \vTwo is larger in the higher \pt range. This is similar to trends observed in ep collisions \cite{zeus:Dec2019_ep,ZEUS:2021qzg}. The increase of \vTwo with \pt  is also present in the simulations although both generators slightly overshoot the data at higher \pt. For \pPb collisions it has been shown that fluctuations in the proton shape can increase \vTwo \cite{Mantysaari:2017cni}. It is noticeable that at a given \pt and \Ntrack, \vTwo is higher for \PhotonP than for \pPb interactions. Tabulated results are provided in the HEPData record for this analysis~\cite{hepdata}.

\section{Summary}

For the first time at the LHC, the study of long-range particle correlations has been extended to photon-proton (\PhotonP) interactions. This study used proton-lead (\pPb) collisions at $\sqrtsNN = 8.16\TeV$ recorded with the CMS detector. The two-particle \VnDelta Fourier coefficients and corresponding single-particle \vTwo azimuthal anisotropies are reported as functions of the multiplicity of charged hadrons (\Ntrack) for two transverse momenta (\pt) ranges. For the \PhotonP sample, the largest observed multiplicity was $\Ntrack\sim35$. The mean \pt of charged particles is smaller in the \PhotonP sample than for \pPb collisions within the same multiplicity range. No evidence for a long-range near-side ridge-like structure was found for either the \PhotonP or hadronic minimum bias \pPb (MB) samples within this multiplicity range. In all \Ntrack and \pt ranges, \VoneDelta is negative, \VtwoDelta is positive with a smaller magnitude than \VoneDelta, and \VthreeDelta is consistent with zero. The magnitudes of both \VoneDelta and \VtwoDelta increase with \pt. This increase has also been seen in electron-proton collisions. At a given \pt and track multiplicity, \vTwo is larger for \PhotonP-enhanced events than for MB \pPb interactions. Predictions from the {\PYTHIA}8 model describe well the \PhotonP data within uncertainties. This suggests the data are dominated by noncollective effects. Within the present experimental sensitivity, no significant collectivity signal is observed.

\begin{acknowledgments}
We congratulate our colleagues in the CERN accelerator departments for the excellent performance of the LHC and thank the technical and administrative staffs at CERN and at other CMS institutes for their contributions to the success of the CMS effort. In addition, we gratefully acknowledge the computing centers and personnel of the Worldwide LHC Computing Grid and other centers for delivering so effectively the computing infrastructure essential to our analyses. Finally, we acknowledge the enduring support for the construction and operation of the LHC, the CMS detector, and the supporting computing infrastructure provided by the following funding agencies: BMBWF and FWF (Austria); FNRS and FWO (Belgium); CNPq, CAPES, FAPERJ, FAPERGS, and FAPESP (Brazil); MES and BNSF (Bulgaria); CERN; CAS, MoST, and NSFC (China); MINCIENCIAS (Colombia); MSES and CSF (Croatia); RIF (Cyprus); SENESCYT (Ecuador); MoER, ERC PUT and ERDF (Estonia); Academy of Finland, MEC, and HIP (Finland); CEA and CNRS/IN2P3 (France); BMBF, DFG, and HGF (Germany); GSRI (Greece); NKFIH (Hungary); DAE and DST (India); IPM (Iran); SFI (Ireland); INFN (Italy); MSIP and NRF (Republic of Korea); MES (Latvia); LAS (Lithuania); MOE and UM (Malaysia); BUAP, CINVESTAV, CONACYT, LNS, SEP, and UASLP-FAI (Mexico); MOS (Montenegro); MBIE (New Zealand); PAEC (Pakistan); MES and NSC (Poland); FCT (Portugal); MESTD (Serbia); MCIN/AEI and PCTI (Spain); MOSTR (Sri Lanka); Swiss Funding Agencies (Switzerland); MST (Taipei); MHESI and NSTDA (Thailand); TUBITAK and TENMAK (Turkey); NASU (Ukraine); STFC (United Kingdom); DOE and NSF (USA).  
     
\hyphenation{Rachada-pisek} Individuals have received support from the Marie-Curie program and the European Research Council and Horizon 2020 Grant, contract Nos.\ 675440, 724704, 752730, 758316, 765710, 824093, 884104, and COST Action CA16108 (European Union); the Leventis Foundation; the Alfred P.\ Sloan Foundation; the Alexander von Humboldt Foundation; the Belgian Federal Science Policy Office; the Fonds pour la Formation \`a la Recherche dans l'Industrie et dans l'Agriculture (FRIA-Belgium); the Agentschap voor Innovatie door Wetenschap en Technologie (IWT-Belgium); the F.R.S.-FNRS and FWO (Belgium) under the ``Excellence of Science -- EOS" -- be.h project n.\ 30820817; the Beijing Municipal Science \& Technology Commission, No. Z191100007219010; the Ministry of Education, Youth and Sports (MEYS) of the Czech Republic; the Hellenic Foundation for Research and Innovation (HFRI), Project Number 2288 (Greece); the Deutsche Forschungsgemeinschaft (DFG), under Germany's Excellence Strategy -- EXC 2121 ``Quantum Universe" -- 390833306, and under project number 400140256 - GRK2497; the Hungarian Academy of Sciences, the New National Excellence Program - \'UNKP, the NKFIH research grants K 124845, K 124850, K 128713, K 128786, K 129058, K 131991, K 133046, K 138136, K 143460, K 143477, 2020-2.2.1-ED-2021-00181, and TKP2021-NKTA-64 (Hungary); the Council of Science and Industrial Research, India; the Latvian Council of Science; the Ministry of Education and Science, project no. 2022/WK/14, and the National Science Center, contracts Opus 2021/41/B/ST2/01369 and 2021/43/B/ST2/01552 (Poland); the Funda\c{c}\~ao para a Ci\^encia e a Tecnologia, grant CEECIND/01334/2018 (Portugal); the National Priorities Research Program by Qatar National Research Fund; MCIN/AEI/10.13039/501100011033, ERDF ``a way of making Europe", and the Programa Estatal de Fomento de la Investigaci{\'o}n Cient{\'i}fica y T{\'e}cnica de Excelencia Mar\'{\i}a de Maeztu, grant MDM-2017-0765 and Programa Severo Ochoa del Principado de Asturias (Spain); the Chulalongkorn Academic into Its 2nd Century Project Advancement Project, and the National Science, Research and Innovation Fund via the Program Management Unit for Human Resources \& Institutional Development, Research and Innovation, grant B05F650021 (Thailand); the Kavli Foundation; the Nvidia Corporation; the SuperMicro Corporation; the Welch Foundation, contract C-1845; and the Weston Havens Foundation (USA).
\end{acknowledgments}
\bibliography{auto_generated}  

\providecommand{\href}[2]{#2}\begingroup\raggedright\begin{thebibliography}{10}%
\makeatletter
\providecommand{\hrefCMSnoop }[0]{\@secondoftwo}%
\makeatother
\providecommand{\doi}{\texttt{doi:}\begingroup \urlstyle{tt}\Url}

\bibitem{cms:ppfirst}
\hrefCMSnoop {}{{CMS Collaboration}, ``Observation of long-range, near-side
  angular correlations in proton-proton collisions at the {LHC}'',} \textit{
  JHEP} \textbf{ 09} (2010) 091,
  \href{http://dx.doi.org/10.1007/JHEP09(2010)091}{\doi{10.1007/JHEP09(2010)091}},
\href{http://www.arXiv.org/abs/1009.4122}{\texttt{arXiv:1009.4122}}.

\bibitem{Aad:2015gqa}
\hrefCMSnoop {}{{ATLAS Collaboration}, ``Observation of long-range elliptic
  azimuthal anisotropies in $\sqrt{s}$ $=$ 13 and 2.76 {\TeV} pp collisions
  with the {ATLAS} detector'',} \textit{ Phys. Rev. Lett.} \textbf{ 116} (2016)
  172301,
  \href{http://dx.doi.org/10.1103/PhysRevLett.116.172301}{\doi{10.1103/PhysRevLett.116.172301}},
  \href{http://www.arXiv.org/abs/1509.04776}{\texttt{arXiv:1509.04776}}.

\bibitem{cms:ppsecond}
\hrefCMSnoop {}{{CMS Collaboration}, ``Measurement of long-range near-side
  two-particle angular correlations in pp collisions at {$\sqrt{s}=13\TeV$}'',}
  \textit{ Phys. Rev. Lett.} \textbf{ 116} (2015) 172302,
  \href{http://dx.doi.org/10.1103/PhysRevLett.116.172302}{\doi{10.1103/PhysRevLett.116.172302}},
\href{http://www.arXiv.org/abs/1510.03068}{\texttt{arXiv:1510.03068}}.

\bibitem{Khachatryan:2016txc}
\hrefCMSnoop {}{{CMS Collaboration}, ``Evidence for collectivity in pp
  collisions at the {LHC}'',} \textit{ Phys. Lett. B} \textbf{ 765} (2017) 193,
  \href{http://dx.doi.org/10.1016/j.physletb.2016.12.009}{\doi{10.1016/j.physletb.2016.12.009}},
  \href{http://www.arXiv.org/abs/1606.06198}{\texttt{arXiv:1606.06198}}.

\bibitem{add_atlas_pp:01}
\hrefCMSnoop {}{{ATLAS Collaboration}, ``Measurement of azimuthal anisotropy of
  muons from charm and bottom hadrons in pp collisions at {$\sqrt{s}=13\TeV$}
  with the {ATLAS} detector'',} \textit{ Phys. Rev. Lett.} \textbf{ 124} (2020)
  082301,
  \href{http://dx.doi.org/10.1103/PhysRevLett.124.082301}{\doi{10.1103/PhysRevLett.124.082301}},
  \href{http://www.arXiv.org/abs/1909.01650}{\texttt{arXiv:1909.01650}}.

\bibitem{cms:pPbfirst}
\hrefCMSnoop {}{{CMS Collaboration}, ``Observation of long-range near-side
  angular correlations in {pPb} collisions at the {LHC}'',} \textit{ Phys.
  Lett. B} \textbf{ 718} (2013) 795,
  \href{http://dx.doi.org/10.1016/j.physletb.2012.11.025}{\doi{10.1016/j.physletb.2012.11.025}},
\href{http://www.arXiv.org/abs/1210.5482}{\texttt{arXiv:1210.5482}}.

\bibitem{add_rhic_pA:01}
\hrefCMSnoop {}{{PHENIX} Collaboration, ``Measurement of long-range angular
  correlations and azimuthal anisotropies in high-multiplicity {p$+$Au}
  collisions at {$\sqrtsNN=200$} {GeV}'',} \textit{ Phys. Rev. C} \textbf{ 95}
  (2017) 034910,
  \href{http://dx.doi.org/10.1103/PhysRevC.95.034910}{\doi{10.1103/PhysRevC.95.034910}},
  \href{http://www.arXiv.org/abs/1609.02894}{\texttt{arXiv:1609.02894}}.

\bibitem{add_rhic_pA:02}
\hrefCMSnoop {}{{PHENIX} Collaboration, ``{Creation of quark-gluon plasma
  droplets with three distinct geometries}'',} \textit{ Nature Phys.} \textbf{
  15} (2019) 214,
  \href{http://dx.doi.org/10.1038/s41567-018-0360-0}{\doi{10.1038/s41567-018-0360-0}},
  \href{http://www.arXiv.org/abs/1805.02973}{\texttt{arXiv:1805.02973}}.

\bibitem{Aad:2012gla}
\hrefCMSnoop {}{{ATLAS Collaboration}, ``Observation of associated near-side
  and away-side long-range correlations in {$\sqrtsNN=5.02\TeV$} proton-lead
  collisions with the {ATLAS} detector'',} \textit{ Phys. Rev. Lett.} \textbf{
  110} (2013) 182302,
  \href{http://dx.doi.org/10.1103/PhysRevLett.110.182302}{\doi{10.1103/PhysRevLett.110.182302}},
\href{http://www.arXiv.org/abs/1212.5198}{\texttt{arXiv:1212.5198}}.

\bibitem{Aad:2013fja}
\hrefCMSnoop {}{{ATLAS Collaboration}, ``Measurement with the {ATLAS} detector
  of multi-particle azimuthal correlations in {p$+$Pb} collisions at
  {$\sqrtsNN=5.02\TeV$}'',} \textit{ Phys. Lett. B} \textbf{ 725} (2013) 60,
  \href{http://dx.doi.org/10.1016/j.physletb.2013.06.057}{\doi{10.1016/j.physletb.2013.06.057}},
\href{http://www.arXiv.org/abs/1303.2084}{\texttt{arXiv:1303.2084}}.

\bibitem{Abelev:2012ola}
\hrefCMSnoop {}{{ALICE Collaboration}, ``Long-range angular correlations on the
  near and away side in {p-Pb} collisions at {$\sqrtsNN=5.02\TeV$}'',} \textit{
  Phys. Lett. B} \textbf{ 719} (2013) 29,
  \href{http://dx.doi.org/10.1016/j.physletb.2013.01.012}{\doi{10.1016/j.physletb.2013.01.012}},
\href{http://www.arXiv.org/abs/1212.2001}{\texttt{arXiv:1212.2001}}.

\bibitem{Aaij:2015qcq}
\hrefCMSnoop {}{{LHCb Collaboration}, ``Measurements of long-range near-side
  angular correlations in {$\sqrtsNN=5\TeV$} proton-lead collisions in the
  forward region'',} \textit{ Phys. Lett. B} \textbf{ 762} (2016) 473,
  \href{http://dx.doi.org/10.1016/j.physletb.2016.09.064}{\doi{10.1016/j.physletb.2016.09.064}},
\href{http://www.arXiv.org/abs/1512.00439}{\texttt{arXiv:1512.00439}}.

\bibitem{ABELEV:2013wsa}
\hrefCMSnoop {}{{ALICE Collaboration}, ``Long-range angular correlations of
  $\pi$, {\PK} and {\Pp} in {pPb} collisions at {$\sqrtsNN=5.02\TeV$}'',}
  \textit{ Phys. Lett. B} \textbf{ 726} (2013) 164,
  \href{http://dx.doi.org/10.1016/j.physletb.2013.08.024}{\doi{10.1016/j.physletb.2013.08.024}},
\href{http://www.arXiv.org/abs/1307.3237}{\texttt{arXiv:1307.3237}}.

\bibitem{Khachatryan:2015waa}
\hrefCMSnoop {}{{CMS Collaboration}, ``Evidence for collective multiparticle
  correlations in {p-Pb} collisions'',} \textit{ Phys. Rev. Lett.} \textbf{
  115} (2015) 012301,
  \href{http://dx.doi.org/10.1103/PhysRevLett.115.012301}{\doi{10.1103/PhysRevLett.115.012301}},
\href{http://www.arXiv.org/abs/1502.05382}{\texttt{arXiv:1502.05382}}.

\bibitem{cms:pPbPbPb_corr_identifiedPar}
\hrefCMSnoop {}{{CMS Collaboration}, ``Long-range two-particle correlations of
  strange hadrons with charged particles in {pPb} and {\PbPb} collisions at
  {LHC} energies'',} \textit{ Phys. Lett. B} \textbf{ 742} (2015) 200,
  \href{http://dx.doi.org/10.1016/j.physletb.2015.01.034}{\doi{10.1016/j.physletb.2015.01.034}},
  \href{http://www.arXiv.org/abs/1409.3392}{\texttt{arXiv:1409.3392}}.

\bibitem{Aaboud:2017acw}
\hrefCMSnoop {}{{ATLAS Collaboration}, ``Measurement of multi-particle
  azimuthal correlations in pp, {p$+$Pb} and low multiplicity {Pb$+$Pb}
  collisions with the {ATLAS} detector'',} \textit{ Eur. Phys. J. C.} \textbf{
  77} (2017) 428,
  \href{http://dx.doi.org/10.1140/epjc/s10052-017-4988-1}{\doi{10.1140/epjc/s10052-017-4988-1}},
\href{http://www.arXiv.org/abs/1705.04176}{\texttt{arXiv:1705.04176}}.

\bibitem{Aaboud:2017blb}
\hrefCMSnoop {}{{ATLAS Collaboration}, ``Measurement of long-range
  multiparticle azimuthal correlations with the subevent cumulant method in pp
  and {p$+$Pb} collisions with the {ATLAS} detector at the {CERN} large hadron
  collider'',} \textit{ Phys. Rev. C} \textbf{ 97} (2018) 024904,
  \href{http://dx.doi.org/10.1103/PhysRevC.97.024904}{\doi{10.1103/PhysRevC.97.024904}},
\href{http://www.arXiv.org/abs/1708.03559}{\texttt{arXiv:1708.03559}}.

\bibitem{PhysRevD.46.229}
\hrefCMSnoop {}{J.-Y. Ollitrault, ``Anisotropy as a signature of transverse
  collective flow'',} \textit{ Phys. Rev. D} \textbf{ 46} (1992) 229,
  \href{http://dx.doi.org/10.1103/PhysRevD.46.229}{\doi{10.1103/PhysRevD.46.229}}.

\bibitem{Heinz:2013th}
\hrefCMSnoop {}{U.~Heinz and R.~Snellings, ``{Collective flow and viscosity in
  relativistic heavy-ion collisions}'',} \textit{ Ann. Rev. Nucl. Part. Sci.}
  \textbf{ 63} (2013) 123,
  \href{http://dx.doi.org/10.1146/annurev-nucl-102212-170540}{\doi{10.1146/annurev-nucl-102212-170540}},
  \href{http://www.arXiv.org/abs/1301.2826}{\texttt{arXiv:1301.2826}}.

\bibitem{Gale:2013da}
\hrefCMSnoop {}{C.~Gale, S.~Jeon, and B.~Schenke, ``Hydrodynamic modeling of
  heavy-ion collisions'',} \textit{ Int. J. Mod. Phys. A} \textbf{ 28} (2013)
  1340011,
  \href{http://dx.doi.org/10.1142/S0217751X13400113}{\doi{10.1142/S0217751X13400113}},
  \href{http://www.arXiv.org/abs/1301.5893}{\texttt{arXiv:1301.5893}}.

\bibitem{Dusling:2015gta}
\hrefCMSnoop {}{K.~Dusling, W.~Li, and B.~Schenke, ``{Novel collective
  phenomena in high-energy proton--proton and proton--nucleus collisions}'',}
  \textit{ Int. J. Mod. Phys. E} \textbf{ 25} (2016) 1630002,
  \href{http://dx.doi.org/10.1142/S0218301316300022}{\doi{10.1142/S0218301316300022}},
  \href{http://www.arXiv.org/abs/1509.07939}{\texttt{arXiv:1509.07939}}.

\bibitem{Nagle:2018eea}
\hrefCMSnoop {}{J.~L. Nagle and J.~Orjuela~Koop, ``A quasiparticle transport
  explanation for collectivity in the smallest of collision systems
  ({$\Pp+\Pp$} and {$\Pe^+\Pe^-$})'',} \textit{ Nucl. Phys. A} \textbf{ 982}
  (2019) 455,
  \href{http://dx.doi.org/10.1016/j.nuclphysa.2018.09.005}{\doi{10.1016/j.nuclphysa.2018.09.005}},
  \href{http://www.arXiv.org/abs/1807.04619}{\texttt{arXiv:1807.04619}}.

\bibitem{add_rhic_AA:01}
\hrefCMSnoop {}{{STAR} Collaboration, ``Elliptic flow in au$+$au collisions at
  {$\sqrtsNN=130$} {GeV}'',} \textit{ Phys. Rev. Lett.} \textbf{ 86} (2001)
  402,
  \href{http://dx.doi.org/10.1103/PhysRevLett.86.402}{\doi{10.1103/PhysRevLett.86.402}},
  \href{http://www.arXiv.org/abs/nucl-ex/0009011}{\texttt{arXiv:nucl-ex/0009011}}.

\bibitem{add_rhic_AA:04}
\hrefCMSnoop {}{{PHOBOS} Collaboration, ``System size dependence of cluster
  properties from two-particle angular correlations in {Cu+Cu} and {Au+Au}
  collisions at {$\sqrtsNN=200$} {GeV}'',} \textit{ Phys. Rev. C} \textbf{ 81}
  (2010) 024904,
  \href{http://dx.doi.org/10.1103/PhysRevC.81.024904}{\doi{10.1103/PhysRevC.81.024904}},
  \href{http://www.arXiv.org/abs/0812.1172}{\texttt{arXiv:0812.1172}}.

\bibitem{cms:PbPbfirst}
\hrefCMSnoop {}{{CMS Collaboration}, ``Long-range and short-range dihadron
  angular correlations in central {\PbPb} collisions at a nucleon-nucleon
  center of mass energy of 2.76 {\TeV}'',} \textit{ JHEP} \textbf{ 10} (2011)
  076,
  \href{http://dx.doi.org/10.1007/JHEP07(2011)076}{\doi{10.1007/JHEP07(2011)076}},
\href{http://www.arXiv.org/abs/1105.2438}{\texttt{arXiv:1105.2438}}.

\bibitem{cms:PbPbsecond}
\hrefCMSnoop {}{{CMS Collaboration}, ``Centrality dependence of dihadron
  correlations and azimuthal anisotropy harmonics in {\PbPb} collisions at
  {$\sqrtsNN=2.76\TeV$}'',} \textit{ Eur. Phys. J. C} \textbf{ 72} (2012)
  10052,
  \href{http://dx.doi.org/10.1140/epjc/s10052-012-2012-3}{\doi{10.1140/epjc/s10052-012-2012-3}},
\href{http://www.arXiv.org/abs/1201.3158}{\texttt{arXiv:1201.3158}}.

\bibitem{ALICE:2011svq}
\hrefCMSnoop {}{{ALICE Collaboration}, ``{Harmonic decomposition of
  two-particle angular correlations in Pb-Pb collisions at $\sqrt{s_{NN}}=$
  2.76 TeV}'',} \textit{ Phys. Lett. B} \textbf{ 708} (2012) 249--264,
  \href{http://dx.doi.org/10.1016/j.physletb.2012.01.060}{\doi{10.1016/j.physletb.2012.01.060}},
  \href{http://www.arXiv.org/abs/1109.2501}{\texttt{arXiv:1109.2501}}.

\bibitem{ATLAS:2012at}
\hrefCMSnoop {}{{ATLAS Collaboration}, ``{Measurement of the azimuthal
  anisotropy for charged particle production in $\sqrt{s_{NN}}=2.76$ TeV
  lead-lead collisions with the ATLAS detector}'',} \textit{ Phys. Rev. C}
  \textbf{ 86} (2012) 014907,
  \href{http://dx.doi.org/10.1103/PhysRevC.86.014907}{\doi{10.1103/PhysRevC.86.014907}},
  \href{http://www.arXiv.org/abs/1203.3087}{\texttt{arXiv:1203.3087}}.

\bibitem{star:AAfirst}
\hrefCMSnoop {}{{STAR} Collaboration, ``{Long range rapidity correlations and
  jet production in high energy nuclear collisions}'',} \textit{ Phys. Rev. C}
  \textbf{ 80} (2009) 064912,
  \href{http://dx.doi.org/10.1103/PhysRevC.80.064912}{\doi{10.1103/PhysRevC.80.064912}},
  \href{http://www.arXiv.org/abs/0909.0191}{\texttt{arXiv:0909.0191}}.

\bibitem{star:AAsecond}
\hrefCMSnoop {}{{STAR} Collaboration, ``{Three-particle coincidence of the long
  range pseudorapidity correlation in high energy nucleus-nucleus
  collisions}'',} \textit{ Phys. Rev. Lett.} \textbf{ 105} (2010) 022301,
  \href{http://dx.doi.org/10.1103/PhysRevLett.105.022301}{\doi{10.1103/PhysRevLett.105.022301}},
  \href{http://www.arXiv.org/abs/0912.3977}{\texttt{arXiv:0912.3977}}.

\bibitem{phenix:AAfirst}
\hrefCMSnoop {}{{PHENIX} Collaboration, ``Formation of dense partonic matter in
  relativistic nucleus-nucleus collisions at {RHIC}: Experimental evaluation by
  the {PHENIX} collaboration'',} \textit{ Nucl. Phys. A} \textbf{ 757} (2005)
  184,
  \href{http://dx.doi.org/10.1016/j.nuclphysa.2005.03.086}{\doi{10.1016/j.nuclphysa.2005.03.086}},
  \href{http://www.arXiv.org/abs/nucl-ex/0410003}{\texttt{arXiv:nucl-ex/0410003}}.

\bibitem{STAR:oct2019}
\hrefCMSnoop {}{{STAR} Collaboration, ``Beam-energy dependence of identified
  two-particle angular correlations in {$\sqrtsNN=7.7-200\GeV$} {Au+Au}
  collisions'',} \textit{ Phys. Rev. C} \textbf{ 101} (2020) 014916,
  \href{http://dx.doi.org/10.1103/PhysRevC.101.014916}{\doi{10.1103/PhysRevC.101.014916}},
  \href{http://www.arXiv.org/abs/1906.09204}{\texttt{arXiv:1906.09204}}.

\bibitem{STAR:oct2018}
\hrefCMSnoop {}{{STAR} Collaboration, ``Correlation measurements between flow
  harmonics in {Au+Au} collisions at {RHIC}'',} \textit{ Phys. Lett. B}
  \textbf{ 783} (2018) 459,
  \href{http://dx.doi.org/10.1016/j.physletb.2018.05.076}{\doi{10.1016/j.physletb.2018.05.076}},
  \href{http://www.arXiv.org/abs/1803.03876}{\texttt{arXiv:1803.03876}}.

\bibitem{Voloshin:1996}
\hrefCMSnoop {}{S.~Voloshin and Y.~Zhang, ``Flow study in relativistic nuclear
  collisions by fourier expansion of azimuthal particle distributions'',}
  \textit{ Phys. Rev. C.} \textbf{ 70} (1996) 665,
  \href{http://dx.doi.org/10.1007/s002880050141}{\doi{10.1007/s002880050141}},
\href{http://www.arXiv.org/abs/hep-ph/9407282}{\texttt{arXiv:hep-ph/9407282}}.

\bibitem{geometry_and_fluctuations_01}
\hrefCMSnoop {}{B.~H. Alver, C.~Gombeaud, M.~Luzum, and J.-Y. Ollitrault,
  ``{Triangular flow in hydrodynamics and transport theory}'',} \textit{ Phys.
  Rev. C} \textbf{ 82} (2010) 034913,
  \href{http://dx.doi.org/10.1103/PhysRevC.82.034913}{\doi{10.1103/PhysRevC.82.034913}},
  \href{http://www.arXiv.org/abs/1007.5469}{\texttt{arXiv:1007.5469}}.

\bibitem{geometry_and_fluctuations_02}
\hrefCMSnoop {}{C.~G. B.~Schenke, S.~Jeon, ``Elliptic and triangular flow in
  event-by-event {$\PD=3+1$} viscous hydrodynamics'',} \textit{ Phys. Rev.
  Lett} \textbf{ 106} (2011) 042301,
  \href{http://dx.doi.org/10.1103/PhysRevLett.106.042301}{\doi{10.1103/PhysRevLett.106.042301}},
\href{http://www.arXiv.org/abs/1009.3244}{\texttt{arXiv:1009.3244}}.

\bibitem{geometry_and_fluctuations_03}
\hrefCMSnoop {}{C.~S. Z.~Qiu and U.~Heinz, ``Hydrodynamic elliptic and
  triangular flow in {Pb-Pb} collisions at {$\sqrtsNN=2.76\TeV$}'',} \textit{
  Phys. Lett. B} \textbf{ 707} (2012) 151,
  \href{http://dx.doi.org/10.1016/j.physletb.2011.12.041}{\doi{10.1016/j.physletb.2011.12.041}},
\href{http://www.arXiv.org/abs/1110.3033}{\texttt{arXiv:1110.3033}}.

\bibitem{elliptical_triangular}
\hrefCMSnoop {}{B.~Alver and G.~Roland, ``Collision geometry fluctuations and
  triangular flow in heavy-ion collisions'',} \textit{ Phys. Rev. C} \textbf{
  81} (2010) 054905,
  \href{http://dx.doi.org/10.1103/PhysRevC.82.039903}{\doi{10.1103/PhysRevC.82.039903}},
  \href{http://www.arXiv.org/abs/1003.0194}{\texttt{arXiv:1003.0194}}.
  [Erratum: Phys. Rev. C 82 (2010) 039903].

\bibitem{cms:april2018}
\hrefCMSnoop {}{{CMS Collaboration}, ``Elliptic flow of charm and strange
  hadrons in high-multiplicity {pPb} collisions at {$\sqrtsNN=8.16\TeV$}'',}
  \textit{ Phys. Rev. Lett.} \textbf{ 121} (2018) 082301,
  \href{http://dx.doi.org/10.1103/PhysRevLett.121.082301}{\doi{10.1103/PhysRevLett.121.082301}},
  \href{http://www.arXiv.org/abs/1804.09767}{\texttt{arXiv:1804.09767}}.

\bibitem{Sirunyan:2018nqr}
\hrefCMSnoop {}{{CMS Collaboration}, ``Centrality and pseudorapidity dependence
  of the transverse energy density in {pPb} collisions at
  {$\sqrtsNN=5.02\TeV$}'',} \textit{ Phys. Rev. C} \textbf{ 100} (2019) 024902,
  \href{http://dx.doi.org/10.1103/PhysRevC.100.024902}{\doi{10.1103/PhysRevC.100.024902}},
  \href{http://www.arXiv.org/abs/1810.05745}{\texttt{arXiv:1810.05745}}.

\bibitem{Adam:2015vsf}
\hrefCMSnoop {}{{ALICE Collaboration}, ``Multi-strange baryon production in
  {pPb} collisions at {$\sqrtsNN=5.02\TeV$}'',} \textit{ Phys. Lett. B}
  \textbf{ 758} (2016) 389,
  \href{http://dx.doi.org/10.1016/j.physletb.2016.05.027}{\doi{10.1016/j.physletb.2016.05.027}},
\href{http://www.arXiv.org/abs/1512.07227}{\texttt{arXiv:1512.07227}}.

\bibitem{add_rhic_AA:02}
\hrefCMSnoop {}{{PHENIX} Collaboration, ``Elliptic flow of identified hadrons
  in {{Au+Au}} collisions at {$\sqrtsNN=200$} {GeV}'',} \textit{ Phys. Rev.
  Lett.} \textbf{ 91} (2003) 182301,
  \href{http://dx.doi.org/10.1103/PhysRevLett.91.182301}{\doi{10.1103/PhysRevLett.91.182301}},
  \href{http://www.arXiv.org/abs/nucl-ex/0305013}{\texttt{arXiv:nucl-ex/0305013}}.

\bibitem{add_rhic_AA:03}
\hrefCMSnoop {}{{STAR} Collaboration, ``Distributions of charged hadrons
  associated with high transverse momentum particles in pp and {Au+Au}
  collisions at {$\sqrtsNN=200$} {GeV}'',} \textit{ Phys. Rev. Lett.} \textbf{
  95} (2005) 152301,
  \href{http://dx.doi.org/10.1103/PhysRevLett.95.152301}{\doi{10.1103/PhysRevLett.95.152301}},
  \href{http://www.arXiv.org/abs/nucl-ex/0501016}{\texttt{arXiv:nucl-ex/0501016}}.

\bibitem{Weller:2017tsr}
\hrefCMSnoop {}{R.~D. Weller and P.~Romatschke, ``One fluid to rule them all:
  viscous hydrodynamic description of event-by-event central p+p, {p+Pb} and
  {Pb+Pb} collisions at {$\sqrt{s}=5.02\TeV$}'',} \textit{ Phys. Lett. B}
  \textbf{ 774} (2017) 351,
  \href{http://dx.doi.org/10.1016/j.physletb.2017.09.077}{\doi{10.1016/j.physletb.2017.09.077}},
  \href{http://www.arXiv.org/abs/1701.07145}{\texttt{arXiv:1701.07145}}.

\bibitem{Bozek:2011if}
\hrefCMSnoop {}{P.~Bozek, ``Collective flow in {p-Pb} and {d-Pb} collisions at
  {\TeV} energies'',} \textit{ Phys. Rev. C} \textbf{ 85} (2012) 014911,
  \href{http://dx.doi.org/10.1103/PhysRevC.85.014911}{\doi{10.1103/PhysRevC.85.014911}},
  \href{http://www.arXiv.org/abs/1112.0915}{\texttt{arXiv:1112.0915}}.

\bibitem{Bozek:2012gr}
\hrefCMSnoop {}{P.~Bozek and W.~Broniowski, ``Correlations from hydrodynamic
  flow in {p-Pb} collisions'',} \textit{ Phys. Lett. B} \textbf{ 718} (2013)
  1557,
  \href{http://dx.doi.org/10.1016/j.physletb.2012.12.051}{\doi{10.1016/j.physletb.2012.12.051}},
  \href{http://www.arXiv.org/abs/1211.0845}{\texttt{arXiv:1211.0845}}.

\bibitem{Bozek:2013uha}
\hrefCMSnoop {}{P.~Bozek and W.~Broniowski, ``{Collective dynamics in
  high-energy proton-nucleus collisions}'',} \textit{ Phys. Rev. C} \textbf{
  88} (2013), no.~1, 014903,
  \href{http://dx.doi.org/10.1103/PhysRevC.88.014903}{\doi{10.1103/PhysRevC.88.014903}},
  \href{http://www.arXiv.org/abs/1304.3044}{\texttt{arXiv:1304.3044}}.

\bibitem{alepCorr:2019}
A.~Badea\hrefCMSnoop {}{ {et~al.}, ``Measurements of two-particle correlations
  in {$\Pe^+\Pe^-$} collisions at {91\GeV} with {ALEPH} archived data'',}
  \textit{ Phys. Rev. Lett.} \textbf{ 123} (2019) 212002,
  \href{http://dx.doi.org/10.1103/PhysRevLett.123.212002}{\doi{10.1103/PhysRevLett.123.212002}},
  \href{http://www.arXiv.org/abs/1906.00489}{\texttt{arXiv:1906.00489}}.

\bibitem{Belle:2022fvl}
\hrefCMSnoop {}{{Belle} Collaboration, ``{Measurement of Two-Particle
  Correlations of Hadrons in e+e- Collisions at Belle}'',} \textit{ Phys. Rev.
  Lett.} \textbf{ 128} (2022), no.~14, 142005,
  \href{http://dx.doi.org/10.1103/PhysRevLett.128.142005}{\doi{10.1103/PhysRevLett.128.142005}},
  \href{http://www.arXiv.org/abs/2201.01694}{\texttt{arXiv:2201.01694}}.

\bibitem{zeus:Dec2019_ep}
\hrefCMSnoop {}{{ZEUS} Collaboration, ``Two-particle azimuthal correlations as
  a probe of collective behaviour in deep inelastic ep scattering at {HERA}'',}
  \textit{ JHEP} \textbf{ 04} (2020) 070,
  \href{http://dx.doi.org/10.1007/JHEP04(2020)070}{\doi{10.1007/JHEP04(2020)070}},
  \href{http://www.arXiv.org/abs/1912.07431}{\texttt{arXiv:1912.07431}}.

\bibitem{ZEUS:2021qzg}
\hrefCMSnoop {}{{ZEUS} Collaboration, ``{Azimuthal correlations in
  photoproduction and deep inelastic $ep$ scattering at HERA}'',} \textit{
  JHEP} \textbf{ 12} (2021) 102,
  \href{http://dx.doi.org/10.1007/JHEP12(2021)102}{\doi{10.1007/JHEP12(2021)102}},
  \href{http://www.arXiv.org/abs/2106.12377}{\texttt{arXiv:2106.12377}}.

\bibitem{ATLAS:2021jhn}
\hrefCMSnoop {}{{ATLAS Collaboration}, ``Two-particle azimuthal correlations in
  photonuclear ultraperipheral {Pb$+$Pb} collisions at {5.02\TeV} with
  {ATLAS}'',} \textit{ Phys. Rev. C} \textbf{ 104} (2021) 014903,
  \href{http://dx.doi.org/10.1103/PhysRevC.104.014903}{\doi{10.1103/PhysRevC.104.014903}},
  \href{http://www.arXiv.org/abs/2101.10771}{\texttt{arXiv:2101.10771}}.

\bibitem{cms:fsq_16_012_01}
\hrefCMSnoop {}{A.~J. Baltz, ``The physics of ultraperipheral collisions at the
  {LHC}'',} \textit{ Phys. Rept.} \textbf{ 458} (2008) 1,
  \href{http://dx.doi.org/10.1016/j.physrep.2007.12.001}{\doi{10.1016/j.physrep.2007.12.001}},
  \href{http://www.arXiv.org/abs/0706.3356}{\texttt{arXiv:0706.3356}}.

\bibitem{cms:fsq_16_012_02}
\hrefCMSnoop {}{C.~F. von Weizs{\"a}cker, ``{Radiation emitted in collisions of
  very fast electrons}'',} \textit{ Z. Phys.} \textbf{ 88} (1934) 612,
\href{http://dx.doi.org/10.1007/BF01333110}{\doi{10.1007/BF01333110}}.

\bibitem{cms:fsq_16_012_03}
\hrefCMSnoop {}{E.~J. Williams, ``{Nature of the high-energy particles of
  penetrating radiation and status of ionization and radiation formulae}'',}
  \textit{ Phys. Rev.} \textbf{ 45} (1934) 729,
\href{http://dx.doi.org/10.1103/PhysRev.45.729}{\doi{10.1103/PhysRev.45.729}}.

\bibitem{cms:fsq_16_012_04}
\hrefCMSnoop {}{E.~Fermi, ``{On the theory of collisions between atoms and
  electrically charged particles}'',} \textit{ Nuovo Cim.} \textbf{ 2} (1925)
  143, \href{http://dx.doi.org/10.1007/BF02961914}{\doi{10.1007/BF02961914}},
  \href{http://www.arXiv.org/abs/hep-th/0205086}{\texttt{arXiv:hep-th/0205086}}.

\bibitem{Sirunyan:2017uyl}
\hrefCMSnoop {}{{CMS Collaboration}, ``Observation of correlated azimuthal
  anisotropy fourier harmonics in pp and p$+${Pb} collisions at the {LHC}'',}
  \textit{ Phys. Rev. Lett.} \textbf{ 120} (2018) 092301,
  \href{http://dx.doi.org/10.1103/PhysRevLett.120.092301}{\doi{10.1103/PhysRevLett.120.092301}},
\href{http://www.arXiv.org/abs/1709.09189}{\texttt{arXiv:1709.09189}}.

\bibitem{Helenius:2019gbd}
\hrefCMSnoop {}{I.~Helenius and C.~O. Rasmussen, ``Hard diffraction in
  photoproduction with {\PYTHIA} 8'',} \textit{ Eur. Phys. J. C} \textbf{ 79}
  (2019), no.~5, 413,
  \href{http://dx.doi.org/10.1140/epjc/s10052-019-6914-1}{\doi{10.1140/epjc/s10052-019-6914-1}},
  \href{http://www.arXiv.org/abs/1901.05261}{\texttt{arXiv:1901.05261}}.

\bibitem{deFavereau:2013fsa}
\hrefCMSnoop {}{{DELPHES 3} Collaboration, ``{DELPHES} 3, a modular framework
  for fast simulation of a generic collider experiment'',} \textit{ JHEP}
  \textbf{ 02} (2014) 057,
  \href{http://dx.doi.org/10.1007/JHEP02(2014)057}{\doi{10.1007/JHEP02(2014)057}},
  \href{http://www.arXiv.org/abs/1307.6346}{\texttt{arXiv:1307.6346}}.

\bibitem{Wang:1991hta}
\hrefCMSnoop {}{X.-N. Wang and M.~Gyulassy, ``{\HIJING}: A monte carlo model
  for multiple jet production in pp, {pA} and {AA} collisions'',} \textit{
  Phys. Rev. D} \textbf{ 44} (1991) 3501,
\href{http://dx.doi.org/10.1103/PhysRevD.44.3501}{\doi{10.1103/PhysRevD.44.3501}}.

\bibitem{Agostinelli:2002hh}
\hrefCMSnoop {}{{GEANT4} Collaboration, ``{\GEANTfour}---a simulation
  toolkit'',} \textit{ Nucl. Instrum. Meth. A} \textbf{ 506} (2003) 250,
  \href{http://dx.doi.org/10.1016/S0168-9002(03)01368-8}{\doi{10.1016/S0168-9002(03)01368-8}}.

\bibitem{Suranyi:2021ssd}
\hrefCMSnoop {}{O.~Sur\'anyi {et~al.}, ``{Performance of the CMS Zero Degree
  Calorimeters in pPb collisions at the LHC}'',} \textit{ JINST} \textbf{ 16}
  (2021), no.~05, P05008,
  \href{http://dx.doi.org/10.1088/1748-0221/16/05/P05008}{\doi{10.1088/1748-0221/16/05/P05008}},
  \href{http://www.arXiv.org/abs/2102.06640}{\texttt{arXiv:2102.06640}}.

\bibitem{Sirunyan:2017ulk}
\hrefCMSnoop {}{{CMS Collaboration}, ``Particle-flow reconstruction and global
  event description with the {CMS} detector'',} \textit{ JINST} \textbf{ 12}
  (2017) P10003,
  \href{http://dx.doi.org/10.1088/1748-0221/12/10/P10003}{\doi{10.1088/1748-0221/12/10/P10003}},
\href{http://www.arXiv.org/abs/1706.04965}{\texttt{arXiv:1706.04965}}.

\bibitem{Chatrchyan:2008zzk}
\hrefCMSnoop {}{{CMS Collaboration}, ``The {CMS} experiment at the {CERN}
  {LHC}'',} \textit{ JINST} \textbf{ 3} (2008) S08004,
\href{http://dx.doi.org/10.1088/1748-0221/3/08/S08004}{\doi{10.1088/1748-0221/3/08/S08004}}.

\bibitem{highqualitytracks}
\hrefCMSnoop {}{{CMS Collaboration}, ``Description and performance of track and
  primary-vertex reconstruction with the {CMS} tracker'',} \textit{ JINST}
  \textbf{ 9} (2014) P10009,
  \href{http://dx.doi.org/10.1088/1748-0221/9/10/P10009}{\doi{10.1088/1748-0221/9/10/P10009}},
  \href{http://www.arXiv.org/abs/1405.6569}{\texttt{arXiv:1405.6569}}.

\bibitem{Khachatryan:2016bia}
\hrefCMSnoop {}{{CMS Collaboration}, ``The {CMS} trigger system'',} \textit{
  JINST} \textbf{ 12} (2017) P01020,
  \href{http://dx.doi.org/10.1088/1748-0221/12/01/P01020}{\doi{10.1088/1748-0221/12/01/P01020}},
  \href{http://www.arXiv.org/abs/1609.02366}{\texttt{arXiv:1609.02366}}.

\bibitem{Sirunyan:2017vpr}
\hrefCMSnoop {}{{CMS Collaboration}, ``Pseudorapidity distributions of charged
  hadrons in proton-lead collisions at $\sqrt{s_{_\mathrm{NN}}}=$ 5.02 and 8.16
  {\TeV}'',} \textit{ JHEP} \textbf{ 01} (2018) 045,
  \href{http://dx.doi.org/10.1007/JHEP01(2018)045}{\doi{10.1007/JHEP01(2018)045}},
  \href{http://www.arXiv.org/abs/1710.09355}{\texttt{arXiv:1710.09355}}.

\bibitem{HIN-18-019}
\hrefCMSnoop {}{{CMS Collaboration}, ``First measurement of the forward
  rapidity gap distribution in {$\mathrm{p}\mathrm{Pb}$} collisions at
  {$\sqrt{s_{_\mathrm{NN}}}=8.16\TeV$}'',}
  \href{http://www.arXiv.org/abs/2301.07630}{\texttt{arXiv:2301.07630}}.

\bibitem{FRG:01}
\hrefCMSnoop {}{{ATLAS Collaboration}, ``Rapidity gap cross sections measured
  with the {ATLAS} detector in pp collisions at {$\sqrt{s}=7\TeV$}'',} \textit{
  Eur. Phys. J. C} \textbf{ 72} (2012) 1926,
  \href{http://dx.doi.org/10.1140/epjc/s10052-012-1926-0}{\doi{10.1140/epjc/s10052-012-1926-0}},
  \href{http://www.arXiv.org/abs/1201.2808}{\texttt{arXiv:1201.2808}}.

\bibitem{FRG:02}
\hrefCMSnoop {}{{CMS Collaboration}, ``Measurement of diffraction dissociation
  cross sections in pp collisions at {$\sqrt{s}=7\TeV$}'',} \textit{ Phys. Rev.
  D} \textbf{ 92} (2015) 012003,
  \href{http://dx.doi.org/10.1103/PhysRevD.92.012003}{\doi{10.1103/PhysRevD.92.012003}},
  \href{http://www.arXiv.org/abs/1503.08689}{\texttt{arXiv:1503.08689}}.

\bibitem{FRG:03}
\hrefCMSnoop {}{J.~D. Bjorken, ``{Rapidity gaps and jets as a new physics
  signature in very high-energy hadron hadron collisions}'',} \textit{ Phys.
  Rev. D} \textbf{ 47} (1993) 101,
  \href{http://dx.doi.org/10.1103/PhysRevD.47.101}{\doi{10.1103/PhysRevD.47.101}}.

\bibitem{cms:pPbthird}
\hrefCMSnoop {}{{CMS Collaboration}, ``Multiplicity and transverse momentum
  dependence of two- and four-particle correlations in p{Pb} and {\PbPb}
  collisions'',} \textit{ Phys. Lett. B} \textbf{ 724} (2013) 213,
  \href{http://dx.doi.org/10.1016/j.physletb.2013.06.028}{\doi{10.1016/j.physletb.2013.06.028}},
  \href{http://www.arXiv.org/abs/1305.0609}{\texttt{arXiv:1305.0609}}.

\bibitem{Mantysaari:2017cni}
\hrefCMSnoop {}{H.~M$\ddot{a}$ntysaari, B.~Schenke, C.~Shen, and P.~Tribedy,
  ``Imprints of fluctuating proton shapes on flow in proton-lead collisions at
  the {LHC}'',} \textit{ Phys. Lett. B} \textbf{ 772} (2017) 681,
  \href{http://dx.doi.org/10.1016/j.physletb.2017.07.038}{\doi{10.1016/j.physletb.2017.07.038}},
  \href{http://www.arXiv.org/abs/1705.03177}{\texttt{arXiv:1705.03177}}.

\bibitem{hepdata}
\hrefCMSnoop {}{``{HEPD}ata record for this analysis'',} 2022.
\newblock
  \href{http://dx.doi.org/10.17182/hepdata.89877}{\doi{10.17182/hepdata.89877}}.

\end{thebibliography}\endgroup
\cleardoublepage \appendix\section{The CMS Collaboration \label{app:collab}}\begin{sloppypar}\hyphenpenalty=5000\widowpenalty=500\clubpenalty=5000
\cmsinstitute{Yerevan Physics Institute, Yerevan, Armenia}
{\tolerance=6000
A.~Tumasyan\cmsAuthorMark{1}\cmsorcid{0009-0000-0684-6742}
\par}
\cmsinstitute{Institut f\"{u}r Hochenergiephysik, Vienna, Austria}
{\tolerance=6000
W.~Adam\cmsorcid{0000-0001-9099-4341}, T.~Bergauer\cmsorcid{0000-0002-5786-0293}, M.~Dragicevic\cmsorcid{0000-0003-1967-6783}, A.~Escalante~Del~Valle\cmsorcid{0000-0002-9702-6359}, R.~Fr\"{u}hwirth\cmsAuthorMark{2}\cmsorcid{0000-0002-0054-3369}, M.~Jeitler\cmsAuthorMark{2}\cmsorcid{0000-0002-5141-9560}, N.~Krammer\cmsorcid{0000-0002-0548-0985}, L.~Lechner\cmsorcid{0000-0002-3065-1141}, D.~Liko\cmsorcid{0000-0002-3380-473X}, I.~Mikulec\cmsorcid{0000-0003-0385-2746}, F.M.~Pitters, N.~Rad, J.~Schieck\cmsAuthorMark{2}\cmsorcid{0000-0002-1058-8093}, R.~Sch\"{o}fbeck\cmsorcid{0000-0002-2332-8784}, M.~Spanring\cmsorcid{0000-0001-6328-7887}, S.~Templ\cmsorcid{0000-0003-3137-5692}, W.~Waltenberger\cmsorcid{0000-0002-6215-7228}, C.-E.~Wulz\cmsAuthorMark{2}\cmsorcid{0000-0001-9226-5812}, M.~Zarucki\cmsorcid{0000-0003-1510-5772}
% ...
\par}
\vskip\cmsinstskip
\dag:~Deceased\\
$^{1}$Also at Yerevan State University, Yerevan, Armenia\\
$^{2}$Also at TU Wien, Vienna, Austria\\
% ...
\end{sloppypar}
\end{document}